# Continuous-wave 6-dB-squeezed light with 2.5-THz-bandwidth from single-mode PPLN waveguide

Takahiro Kashiwazaki,[1,a)] Naoto Takanashi,[2] Taichi Yamashima,[2] Takushi Kazama,[1] Koji Enbutsu,[1] Ryoichi Kasahara,[1] Takeshi Umeki,[1] and Akira Furusawa[2,b)]

[1] *NTT Device Technology Labs, NTT Corporation, 3-1, Morinosato Wakamiya, Atsugi, Kanagawa, 243-0198, Japan*

[2] *Department of Applied Physics, School of Engineering, The University of Tokyo, 7-3-1 Hongo, Bunkyo-ku, Tokyo, 113-8656, Japan*

[a)] Electronic mail: takahiro.kashiwazaki.dy@hco.ntt.co.jp
[b)] Electronic mail: akiraf@ap.t.u-tokyo.ac.jp

# Continuous-wave 6-dB-squeezed light with 2.5-THz-bandwidth from single-mode PPLN waveguide

Takahiro Kashiwazaki,[1),a)] Naoto Takanashi,[2)] Taichi Yamashima,[2)] Takushi Kazama,[1)] Koji Enbutsu,[1)] Ryoichi Kasahara,[1)] Takeshi Umeki,[1)] and Akira Furusawa[2),b)]

[1)] *NTT Device Technology Labs, NTT Corporation, 3-1, Morinosato Wakamiya, Atsugi, Kanagawa, 243-0198, Japan*

[2)] *Department of Applied Physics, School of Engineering, The University of Tokyo, 7-3-1 Hongo, Bunkyo-ku, Tokyo, 113-8656, Japan*

[a)] Electronic mail: takahiro.kashiwazaki.dy@hco.ntt.co.jp
[b)] Electronic mail: akiraf@ap.t.u-tokyo.ac.jp

**ABSTRACT**

Terahertz (THz)-bandwidth continuous-wave (CW) squeezed light is essential for integrating quantum processors with time-domain multiplexing (TDM) by using optical delay line interferometers. Here, we utilize a single-pass optical parametric amplifier (OPA) based on a single-spatial-mode periodically poled ZnO:LiNbO$_3$ waveguide, which is directly bonded onto a LiTaO$_3$ substrate. The single-pass OPA allows THz bandwidth, and the absence of higher-order spatial modes in the single-spatial-mode structure helps to avoid degradation of squeezing. In addition, the directly bonded ZnO-doped waveguide has durability for high-power pump and shows small photorefractive damage. Using this waveguide, we observe CW 6.3-dB squeezing at 20-MHz sideband by balanced homodyne detection. This is the first realization of CW squeezing with a single-pass OPA at a level exceeding 4.5 dB, which is required for the generation of a two-dimensional cluster state. Furthermore, the squeezed light shows 2.5-THz spectral bandwidth. The squeezed light will lead to the development of a high-speed on-chip quantum processor using TDM with a centimeter-order optical delay line.

## I. INTRODUCTION

Quantum information processing (QIP) with quadrature amplitudes of light is promising for realizing deterministic, universal, and fault-tolerant operation.[1,2] In particular, continuous-wave (CW) light enables efficient implementation of large-scale QIP with time-domain multiplexing (TDM) by using delay line interferometers.[3] This is because CW light provides full utilization of processing time, unlike pulsed light, which wastes most of the time due to its low duty ratio. Applying TDM for QIP enables us, in principle, to generate unlimited scales of quantum entanglement in finite space.[1,4] Recently, large-scale quantum entanglement of over one-million states[5] and two-dimensional (2D) cluster states[6] have been successfully demonstrated by using TDM.

In the TDM scheme, generation of broadband CW squeezed light is essential for speeding up and downsizing quantum processors. A wide bandwidth is desired for high-speed quantum processing because the maximum clock rate of a quantum processor is limited by the bandwidth of the quantum resource, namely squeezed light. At the same time, the wide bandwidth contributes to downsizing the system. Terahertz (THz)-order broadband squeezed light allows us to define micrometer-order wave packet modes, which require a delay line with only a 3-cm optical length for a cluster state with one-hundred entangled modes for quantum computing. The downsizing enables us to integrate the processors into a silica-based optical chip.[7,8]

For a quantum processor, not only broadband light but also high-level squeezing is required to obtain the "quantum" effect, such as at least 3 dB for entanglement swapping,[8,9] and 4.5 dB for the generation of 2D cluster states.[6] Since the first observation of squeezed light in 1985,[10] various approaches have been used to achieve high-level squeezing.[11] However, over-4.5-dB squeezing, which is required for the generation of 2D cluster states, has not yet been achieved for CW light with a THz bandwidth.

To obtain high-level CW squeezed light, an optical parametric oscillator (OPO) is usually used with a second-ordered nonlinear optical crystal inserted into an optical cavity, which enhances the optical parametric process. The CW squeezing levels of 7.2, 9.0, and 15.3 dB were achieved in 2006,[12] 2007,[13] and 2016,[14] respectively. Although these experimental results have demonstrated the high squeezing level required for practical application, the squeezing bandwidth is limited by the cavity-based configuration itself. Even with a monolithically integrated OPO, the bandwidth of the squeezing light is up to 2.5 GHz.[15]

The use of a single-pass traveling-wave optical parametric amplifier (OPA) is an attractive way to obtain broadband squeezed light. Squeezing with THz bandwidth has been demonstrated thanks to the wide bandwidth of OPAs, which is, in principle, limited only by the transparency or phase matching conditions of the nonlinear medium itself.[16] One way to achieve a high squeezing level from a single-pass OPA is to pump it with pulsed light, which instantaneously provides large electric-field amplitude.[17,18] The drawback of using pulsed light lies in the requirement of shaping a local oscillator (LO) spatiotemporally to match with the generated squeezed light.[19,20] In particular, spatial mode matching is difficult because the phase front of squeezed light from a single-pass OPA is distorted by gain-induced diffraction (GID) in a nonlinear crystal with high parametric gain.[21,22] With spatiotemporally matched LO pulses, 5.8-dB squeezing was obtained from a type II $KTiOPO_4$ (KTP) crystal in 1994,[21] which stood as the world record with a single-pass OPA system for over 20 years. In the CW scheme, OPAs require an alternative way to enhance nonlinearity without a cavity configuration and high-peak-power pumping scheme. In 1995, 0.7-dB squeezing was demonstrated using a waveguide device,[23] which provides high nonlinearity by finely confining light in a small core through its long interaction length.[24,25] Waveguide devices have also yielded a reduction of the GID effect thanks to good spatial mode overlapping.[26,27] In addition, there is no temporal mode mismatching in the CW-pumping scheme. So far, 2.0-dB[28] and 2.2-dB[25] squeezing have been demonstrated by using CW-pumped OPAs based on highly efficient periodically poled $LiNbO_3$ (PPLN) and KTP waveguides, respectively.

To increase the squeezing level with a waveguide, we have to overcome drawbacks arising from the waveguide configuration and the CW-pumping scheme. In a waveguide, high-order spatial modes degrade the measurable squeezing level for the fundamental mode because the waveguide's output can be a mixture of a squeezed quadrature in the fundamental mode and anti-squeezed quadratures in higher-order modes. In the CW-pumping scheme, the issue is the resistance for high-power pumping. In nonlinear crystal waveguides, pump-induced optical scattering, known as the photorefractive effect,[29,30] is inevitable because the average power required for CW-pumping OPA systems is over hundred times higher than that for pulsed-pumping ones.

In this paper, we report the detection of 6.3-dB CW squeezed light at 20-MHz sidebands using a single-mode PPLN waveguide directly bonded onto a $LiTaO_3$ substrate. The bandwidth of the squeezed light is estimated to be 2.5 THz from parametric fluorescence spectrum. The single-mode structure avoids contamination of the squeezed quadrature by the anti-squeezed quadratures of higher-order spatial-modes. The core material of lithium niobate was doped with ZnO. The directly bonded ZnO-doped PPLN waveguide has durability for high-power pump and shows small photorefractive damage with an over-100-mW CW pump at 20°C. The squeezing level is high enough for the generation of 2D cluster states, which require squeezed light of over 4.5 dB as a resource. Furthermore, the result breaks the records of both single-pass CW 2.2-dB squeezing achieved with a periodically-poled KTP waveguide in 2009[25] and single-pass pulsed 5.8-dB squeezing with a bulk KTP crystal in 1994.[21]

## II. DEVICE DESIGN AND FABRICATION

To detect higher squeezing levels from a waveguide OPA, it is important to realize a highly efficient optical parametric process, low-loss detection, and high-power pump injection as described in[23]

$$R_\pm = 1 - \eta + \eta \times \exp(\pm 2\sqrt{aP}), \quad (1)$$

where $R_-$ is the squeezing level, $R_+$ is an anti-squeezing level, $\eta$ is total effective detection efficiency for the generated squeezed light, $a$ is the second-harmonic (SH) conversion efficiency of the waveguide OPA, and $P$ is pump power in the waveguide. Recently, higher values of nonlinear coefficient $a$ have been obtained thanks to an improvements in waveguide fabrication technologies.[31] On the other hand, $P$ and $\eta$ are remaining issues. Acceptable pump power of the waveguide is limited by pump-induced phenomena such as photorefractive damage.[29,30] Therefore, we need to improve the durability of the core material for high-power pump light so that the pump power can be increased with a low photorefractive effect. Regarding optical loss, $\eta$ can be divided into several factors as expressed by

$$\eta = (1 - L_{WG})(1 - L_{HD}). \quad (2)$$

Here, $L_{WG}$ and $L_{HD}$ are effective transmission loss of the waveguide for squeezed light and measurement loss of homodyne detection, respectively.

$L_{WG}$ includes intrinsic loss of optical materials, extrinsic loss derived from structural imperfections in the waveguide, and pump-induced loss of the nonlinear optical crystals. To reduce pump-induced loss, a highly durable nonlinear crystal for a high-power pump should be used. In addition, to maintain durability for high-power pump, it is important to avoid deteriorating the crystal quality during waveguide fabrication. As described before, when these requirements are met, we can not only suppress the loss but use a high-power pump as well, which is desired for the generation of high-level squeezed light.

The homodyne detection loss, $L_{HD}$, is derived from the spatiotemporal mode-mismatch with an LO, photo-electric conversion loss of photodiodes, and electrical circuit noise equivalent to optical loss. The circuit noise can be made negligibly small by using sufficient LO power. Detector's quantum efficiency of up to 99% has been achieved thanks to recent improvements in fiber optic components for telecommunications.[32] However, mode-mismatch is still a problem in single-pass OPA squeezing. In CW squeezing, the problem is spatial mode-mismatch rather than temporal mode-mismatch. The difficulty of spatial mode-match has often been addressed by considering GID,[21] which is caused by a difference in the spatial-modal-shape between the pump and squeezed light in a nonlinear bulk crystal with a single-pass amplification scheme. The phase front of the squeezed light is distorted by GID in a high-parametric-gain regime. It has been reported that using a waveguide can suppress GID;[26,27] however, a similar problem is thought to occur in a multi-mode waveguide. The generated squeezed light in a multi-mode waveguide contains various spatial modes, and the phase and amplitude of each mode have a complex relationship. Therefore, it is difficult to shape the LO to have completely the same optical phase-amplitude distribution as squeezed light, even though mode-shaping techniques are used for LOs. For this reason, the measured squeezing quadrature is usually contaminated by anti-squeezing quadratures from higher-order spatial modes. Here, one of the solutions is to use a single-mode waveguide, which can completely eliminate the contamination problem of a spatially multi-mode OPA.

For single-mode propagation of squeezed light, we calculated a modal dispersion curve of a waveguide by the finite-difference method (Optiwave System Inc., OptiBPM 12). Here, we assumed ZnO-doped PPLN as a core material for its high second-ordered nonlinearity and broadband transparency. ZnO is used as a dopant to suppress the photorefractive effect of lithium niobate.[33] In the calculation, we assumed a core thickness of 5.0 µm, as shown in Fig. 1(a). The shape of the cross-section of the waveguide is trapezoidal, which is inevitable due to redeposition of lithium niobate in the dry etching process.[34] Our waveguide has a side wall angle of 73.5° on both sides. Figure 1(b) shows effective refractive indices of three modes for vertically polarized 1550-nm light as a function of waveguide-top width. Insets show modal amplitude distributions for fundamental, second-order, and third-order modes with the waveguide-top widths of 4.0, 6.5, and 11.0 µm, respectively. The higher ordered modes are allowed to exist in the region with over a 6.5-µm top width. Generally, a large size core is tolerant against fabrication errors, such as roughness of the waveguide core's side walls. Thus, we decided to make the waveguide top width about 6.0 µm, which is the maximum width for the single-mode condition at the wavelength of 1550 nm.

We fabricated a ZnO-doped PPLN waveguide by direct bonding and dry etching.[35] The waveguide length was 45 mm. The quasi-phase matching pattern was fabricated by the electrical poling process with the poling period of about 18 µm. Both the input and output end-faces of the waveguide were mechanically polished and covered with an anti-reflection (AR) coating for both 1550-nm and 775-nm light. Figure 1(c) shows cross-sectional and perspective views of the fabricated PPLN waveguide observed by scanning electron-beam microscopy. The waveguide was directly bonded onto a $LiTaO_3$ wafer, whose thermal expansion coefficient is close to that of lithium niobate. The bonded waveguides showed high durability for high-power second-harmonic (SH) pump light and high-efficiency optical parametric conversion,[31,36] compared to a waveguide fabricated by the titanium diffusion method.[37] The ion diffusion process causes a defect in the core crystal of the waveguide and deteriorates the durability for high-power pump light.[38] Figure 1(d) shows the phase-matching curve of the fabricated PPLN waveguide at 45°C. The SH-conversion efficiency was estimated by the powers of output SH light and input 3-mW fundamental light. At the center wavelength of 1552 nm, our waveguide shows 1160% $W^{-1}$ of SH conversion efficiency.

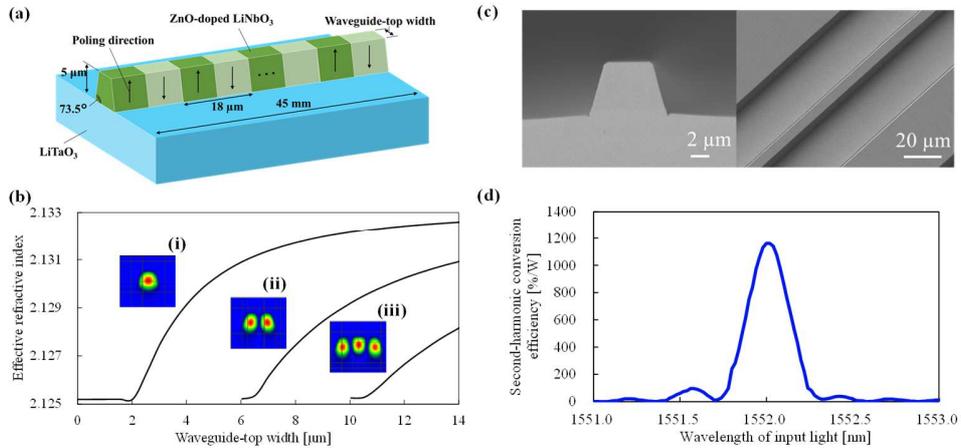

FIG. 1. (a) Schematic view of fabricated PPLN waveguide. (b) Modal dispersion curve for PPLN waveguide calculated by finite differential method. Effective refractive indices for three modes as a function of a waveguide top width are shown. (i) First-order mode, namely the fundamental mode, with a waveguide-top width of 4.0 µm. (ii) Second-order mode with 6.5 µm. (iii) Third-order mode with 11.0 µm. Calculated shape for each mode is also shown. The waveguide structure was assumed to be fabricated by dry etching method. Thickness of core was 5.0 µm, and wall angles of waveguide were 73.5°. (c) Cross-sectional and perspective views of the PPLN waveguide taken with scanning electron-beam microscopy. The PPLN waveguide is directly bonded onto a lithium tantalate substrate. (d) Second-harmonic conversion efficiency of fabricated PPLN waveguide. A power of input light was 3 mW.

## III. EXPERIMENTAL RESULTS

### A. Squeezing-level measurement

Figure 2 shows the experimental setup for detection of a squeezed vacuum from a fabricated PPLN waveguide (PPLN-WG1). CW light from a fiber laser at a wavelength of 1550.0 nm (NKT Photonics, Koheras BOOSTIK) was distributed into two optical fibers by an optical fiber beamsplitter (Thorlabs, PN1550R2A1). Both beams passed through variable optical attenuators (VOA1 and VOA2) (Thorlabs, VOA50PM-APC). One beam was injected into a frequency doubler based on a PPLN waveguide (NTT Electronics, WH-0776-000-F-B-C). A frequency-doubled beam was used as a pump beam for PPLN-WG1 and generated squeezed light according to the optical parametric process. The temperature of PPLN-WG1 was about 20°C, which was controlled by a Peltier device to keep phase-matching condition for 1550 nm. The squeezed vacuum was separated from the SH pump beam by a dichroic mirror (DM). The other output from the fiber beamsplitter was used as an LO. For homodyne detection, the squeezed vacuum gets interfered with the LO at the 50/50 beamsplitter. A different PPLN waveguide (PPLN-WG2), which has the same structure as PPLN-WG1, was used as a beam shaper for the LO to get high spatial mode-matching in the homodyne detection. Here, the visibility was about 0.98 in the experimental setup. We emphasize that the single-mode structures of PPLN-WG1 and PPLN-WG2 largely contributed to this high visibility. The phase-matched wavelength of PPLN-WG2 was intentionally detuned by controlling waveguide temperature to suppress second-order nonlinear processes, such as second-harmonic generation. The phase of the LO beam was scanned by a phase modulator (Thorlabs, LN65-10-P-F-F-BNL) driven by a 500-mHz triangle wave signal. The homodyne detector consisted of specially ordered InGaAs photodiodes (Laser Components, IGHQEX0100-1550-10-1.0-SPAR-TH-40) and an AC-coupled trans-impedance amplifier, which was also used in Ref. 32. The quantum efficiency of the photodiode is about 99% at 1550 nm. The output electric signal was measured by an electrical spectrum analyzer (ESA) (Agilent, E4401B).

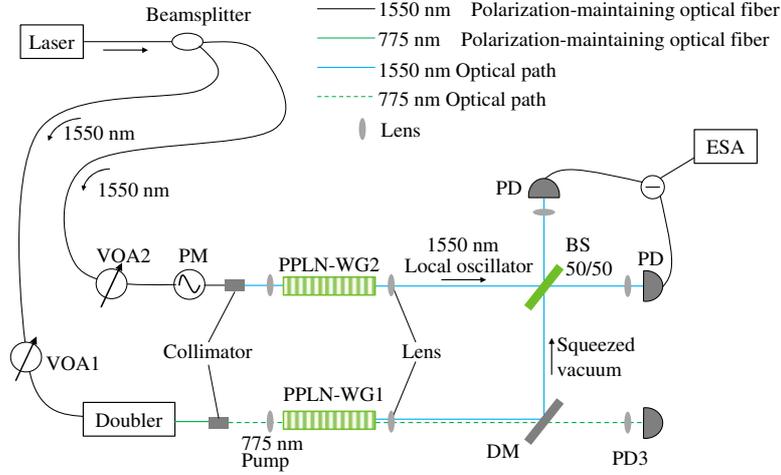

FIG. 2. Schematic diagram of the experimental setup for detection of squeezed light from a PPLN waveguide. BS: 50/50 beamsplitter. DM: Dichroic mirror, which reflects squeezed light and through pump light. PD: Photo detector. VOA: Variable optical attenuator. PM: Optical phase modulator for phase scanning of homodyne detection. PPLN-WG1: PPLN waveguide for generation of squeezed vacuum. PPLN-WG2: PPLN waveguide for mode shaping. ESA: Electrical spectrum analyzer.

Figure 3(a) shows powers of squeezed vacuum noise and squeezed noise detected by a zero span measurement with the spectrum analyzer at the frequency of 20 MHz. The pump power injected into the waveguide was 304 mW, which was estimated by the power detected at PD3. The resolution bandwidth and video bandwidth were 5 MHz and 3 kHz, respectively. Detected noise was amplified and de-amplified by scanning the LO phase. Since the scan signal for the modulator was a triangle wave, the detected noise shows symmetrical shape centered at the sweep time of about 0.25 s. The measured squeezed and anti-squeezed noise levels are -6.3±0.1 and 14.7±0.1 dB, respectively. The shot noise level was measured by blocking the optical path soon after the PPLN-WG1. The power of output LO beam from PPLN-WG2 was 8.9 mW. The measurement error was defined as the standard variation of shot noise of 0.1 dB. The circuit noise level of homodyne measurement was about -50 dBm, which is negligibly small compared to the shot noise level of about -30 dBm.

Figure 3(b) shows the squeezed and anti-squeezed noise levels as functions of pump power in PPLN-WG1. The pump power was controlled by VOA1. The squeezing levels and anti-squeezing levels were measured at the frequency of 20 MHz. The anti-squeezed noise levels increased with increasing pump power. On the other hand, the squeezed noise levels saturated at about -6 dB due to the effective loss for squeezed light. This asymmetrical shape was caused by contamination of the vacuum noise derived from the optical loss for squeezed light. In this experiment, the LO phase scanning was much slower than the averaging time of the spectrum analyzer, and we could ignore waveform blunting due to the measurement setting. The solid fitting curves are drawn according to Eq. (1), and $\eta$ and $a$ are fitted to 0.79 and 1210% W$^{-1}$, respectively. Considering the photodiode's quantum efficiency of 0.99, optical transmittance through the beamsplitter and the mirrors of 0.97, and the visibility of 0.98, the total transmittance for the output squeezed vacuum is 0.94, which equivalent to $L_{HD}$ of 0.06. Therefore, the estimated effective loss in the PPLN waveguide, $L_{WG}$, is 0.16.

The anti-squeezed noise level exponentially increases as the SH- pump power increases, even when it reaches over 100 mW. Furthermore, the estimated SH conversion efficiency of 1210% $W^{-1}$ is consistent with the obtained value of 1160% $W^{-1}$, which was estimated from the measurement of SH conversion efficiency with a 3-mW 1.5-µm-wavelength pump light shown in Fig. 1(d). These results indicate that our waveguide provides a highly efficient optical parametric process without degradation due to pump-induced loss, such as that due to photorefractive damage. In addition, our PPLN waveguide works at 20°C, indicating that the directly bonded ZnO-doped PPLN waveguide has high durability for high-power pumping compared to other nonlinear waveguides that must be operated at high temperature to suppressing photorefractive damage.[33,39] A higher squeezing value could be obtained by improving the fabrication process to reduce $L_{WG}$, especially the optical propagation loss due to structural imperfections. Realization of higher SH-conversion efficiency would be also important, for example, by using a smaller core waveguide.[40]

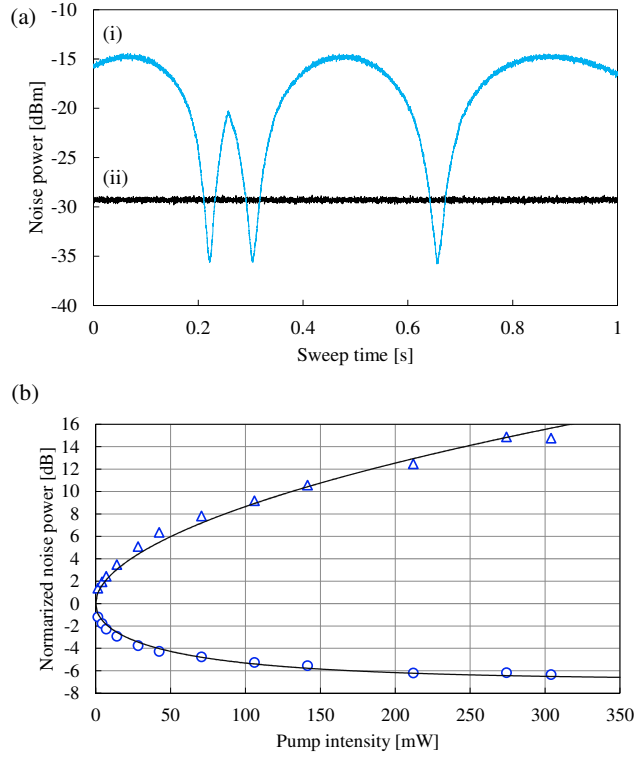

FIG. 3. (a) Raw data of noise power detected by homodyne detection as a function of sweep time with the phase scanning of the LO beam. (i) Noise of squeezed vacuum. (ii) Shot noise. The scan signal for the modulator is a 500-mHz triangle wave. Measurement frequency is 20 MHz. Resolution bandwidth is 5 MHz. Video bandwidth is 3 kHz. The intensity of the pump beam is 304 mW, measured by PD3. The circuit noise level is lower than the shot noise level by more than 20 dB. (b) Squeezing and anti-squeezing levels as functions of pump intensities detected by PD3. The levels are normalized by the intensity of shot noise. The measured noise powers were obtained at a fixed analysis frequency of 20 MHz. Resolution bandwidth is 5 MHz. The solid line is a fitting curve. The pump power in the PPLN-WG1 was estimated from the detected power at PD3.

## B. Bandwidth measurement of squeezed light

To estimate the bandwidth of the squeezed light, we measured squeezing levels at various frequencies with the same experimental setup as that shown in Fig. 2. Figure 4 shows observed noise power at each frequency from 10 MHz to 1 GHz. The resolution bandwidth was 5 MHz. The shot noise and the circuit noise spectra are also shown. The circuit noise was detected by blocking both ports of the homodyne detector. At up to 250 MHz, the squeezing levels stayed over 4.5 dB, which is the value required for generating a 2D cluster state. Above 300 MHz, the squeezing levels decreased with decreasing electrical circuit noise. This means that the measured squeezing level in the high-frequency range was limited by the measurement capability of the homodyne system, such as the bandwidth of the photodiodes.

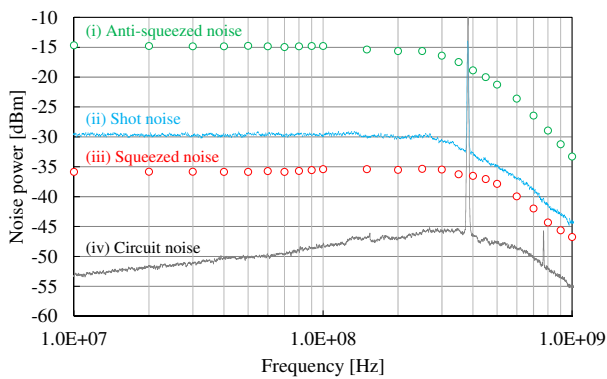

FIG. 4. Measured noise powers detected at various frequencies. (i) Anti-squeezed noise. (ii) Shot noise. (iii) Squeezed noise. (iv) Circuit noise. Resolution bandwidth is 5 MHz. The circuit noise was detected by blocking both ports of a homodyne detector.

In order to estimate an actual squeezing bandwidth, we measured the optical spectrum of the squeezed light from a fabricated PPLN waveguide (PPLN-WG1). Fig. 5(a) shows the experimental setup. Optical components in the setup were the same as those used in the measurement for squeezing levels (Fig. 2). Squeezed light from PPLN-WG1 was separated from a SH beam by a dichroic mirror (DM) and injected into an optical spectrum analyzer (Anritsu, MS9710C). Fig. 5(b) shows the spectra at various pump powers. The optical resolution and video bandwidth were 50 pm and 100 Hz, respectively. The squeezed light, consisting of pairs of signal and idler photons, has an almost symmetrical shape with a center frequency of 193.4 THz, corresponding to a center wavelength of 1550 nm. The generated squeezed light has 2.5-THz bandwidth as its half width at half maximum, which directly represents the squeezing bandwidth.[41] To maintain and utilize this broad bandwidth for actual applications, it is important to prepare optical components with low dispersion for squeezed light or to use inverse dispersion media for dispersion compensation. Furthermore, using integrated high-speed electro-optic devices[42] and using a broadband detection technique with optical parametric amplifier[43] will lead to realize high-speed on-chip quantum processors.

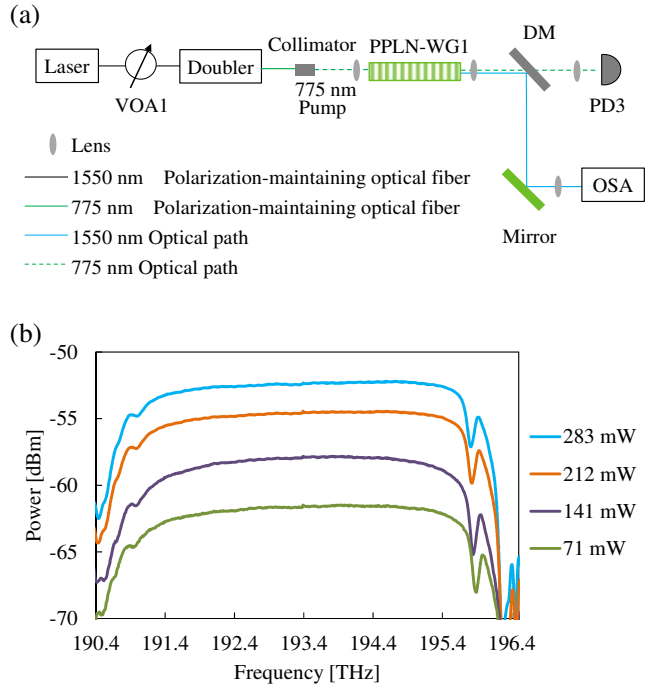

FIG. 5. (a) Schematic diagram of the experimental setup for detection of squeezed light from a PPLN waveguide by an optical spectrum analyzer. DM: Dichroic mirror, which reflects squeezed light and through pump light. PD3: Photodetector for pump light. VOA: Variable optical attenuator. PPLN-WG1: PPLN waveguide for generation of squeezed light. OSA: Optical spectrum analyzer. (b) Spectrum of squeezed light with various powers of pump light, detected by an optical spectrum analyzer. The pump power was estimated from the power detected at PD3. Optical resolution and video bandwidth are 50 pm and 100 Hz, respectively.

## IV. CONCLUSION

In summary, we detected CW 6.3-dB squeezed light at 20 MHz sidebands by using a single-mode PPLN waveguide directly bonded on to a $LiTaO_3$ substrate. Squeezing at over 4.5 dB was measured up to a frequency of 250 MHz by balanced homodyne measurement. The actual bandwidth of the squeezed light is estimated to be 2.5 THz by measurement of its optical spectrum. The single-mode structure avoids contamination of the squeezed quadrature by the anti-squeezed quadrature of higher-order spatial modes. The directly bonded ZnO-doped PPLN has durability for high-power pump and shows small photorefractive damage with a CW pump of over 100 mW at 20°C. The squeezed light with a bandwidth exceeding 1 THz will enable us to define micrometer-order wave-packet modes, which requires only a delay line with a 3-cm optical length in differential interferometers for hundreds of entangled modes in QIP with TDM. This means that the squeezed light will open the possibility of large-scale QIP with a small optical system, such as with an integrated optical chip. We believe that our results will lead to the development of a high-speed optical quantum processor with a large-scale cluster state.

**ACKNOWLEDGEMENT**

This work was supported by Core Research for Evolutional Science and Technology (CREST) (JPMJCR15N5) of Japan Science and Technology Agency (JST), KAKENHI (18H05297) of Japan Society for the Promotion of Science (JSPS), APLS of Ministry of Education, Culture, Sports, Science and Technology (MEXT), and The University of Tokyo Foundation. The author acknowledges Asuka Inoue and Yoshiki Nishida for their useful comments on the manuscript.